\documentclass[preprint,one-column,nofootinbib,english]{revtex4-1}
\usepackage{amsthm,amsmath,amsfonts,dsfont,upgreek}    
\usepackage{amssymb,epsfig,setspace}
\usepackage{graphicx}
\usepackage{babel}
\usepackage{verbatim}

\newcommand{\Dth}{\underset{h}{\Delta}}
\newcommand{\sint}{\int\!\!\!\!\!\!\!\!\!\;\sum}

\begin{document}

\hyphenation{ano-ther ge-ne-ra-te dif-fe-rent know-le-d-ge po-ly-no-mi-al}
\hyphenation{me-di-um  or-tho-go-nal as-su-ming pri-mi-ti-ve pe-ri-o-di-ci-ty}
\hyphenation{mul-ti-p-le-sca-t-te-ri-ng i-te-ra-ti-ng e-q-ua-ti-on}
\hyphenation{wa-ves di-men-si-o-nal ge-ne-ral the-o-ry sca-t-te-ri-ng}
\hyphenation{di-f-fe-r-ent tra-je-c-to-ries e-le-c-tro-ma-g-ne-tic pho-to-nic}
\hyphenation{Ray-le-i-gh di-n-ger Kra-jew-ska Wal-czak Ham-bur-ger Ad-di-ti-o-nal-ly}
\hyphenation{Kon-ver-genz-the-o-rie ori-gi-nal in-vi-si-b-le cha-rac-te-ri-zed}
\hyphenation{Ne-ver-the-less sa-tu-ra-te Ene-r-gy sa-ti-s-fy le-vels re-s-pec-ti-ve pro-pe-r-ty}
\hyphenation{dif-fe-rent no-men-cla-tu-re re-gar-ding}

\title{On uniqueness of Heine-Stieltjes polynomials for second order
finite-difference equations}

\author{Alexander Moroz} 

\affiliation{Wave-scattering.com} 
 
\begin{abstract}
\noindent 
A second order finite-difference equation has two linearly independent solutions.
It is shown here that, like in the continuous case, at most one of the two can be a polynomial solution. 
The uniqueness in the classical continuous Heine-Stieltjes theory 
is shown to hold under broader hypotheses than usually presented. 
A difference between regularity condition and uniqueness is emphasized.
Consistency of our uniqueness results is also checked against one of the Shapiro problems.
An intrinsic relation between the Heine-Stieltjes problem and the 
{\em discrete} Bethe Ansatz equations allows one to immediately extend 
the uniqueness result from the former to the latter.
The results have implications for nondegeneracy of polynomial
solutions of physical models. 
\end{abstract}


\maketitle

\section{Introduction}
\label{sc:intr}
Let $A(x)$ and $B(x)$ be given polynomials of degrees
$m+1$ and $m$, respectively. The subject of the classical 
Heine-Stieltjes theory is to determine 
a polynomial $V(x)$ of degree $m-1$ such that the second-order 
differential equation
\begin{equation}
A(x)y''+2B(x)y'+V(x)y=0
\label{s6.8.1}
\end{equation}
has a solution which is a polynomial of a preassigned degree $n$ \cite{Heine1878,Stl,Vlk,Bch,Sz}.
Assume with Stieltjes \cite{Stl} that $A(x)$ has {\em real} unequal roots,
\begin{equation}
A(x) = (x- a_0)(x- a_1)\ldots (x- a_m),\qquad a_0< a_1 < \ldots  < a_m,
\label{s6.81.1} 
\end{equation}
and
\begin{equation}
{B(x)\over A(x)} = {\rho_0\over x- a_0}+ {\rho_1\over x- a_1}
   +\ldots {\rho_m\over x- a_m},\qquad \rho_\nu>0, ~~~~ \nu=0,1,2,\ldots ,m.
\label{s6.81.2}
\end{equation}
This is equivalent to the assumption that the zeros of $A(x)$ {\em alternate} with
those of $B(x)$ and that the leading order coefficients of $A(x)$ and $B(x)$ have the
{\em same} sign.
Under the above conditions, the basic properties of polynomial solutions 
are \cite{Stl,Bch,Sz,Shp}:
\begin{itemize}

\item there are exactly
\begin{equation}
\sigma_{nm} = {n+m-1 \choose n}
\nonumber 
\end{equation}
polynomials $V(x)$, which are called {\em van Vleck} polynomials \cite{Shp,MMM,PBD,ML}.

\item Eq. (\ref{s6.8.1}) cannot have two polynomial
solutions linearly {\em independent} of each other.

\item If $y$ is a polynomial solution, $y\not\equiv 0$, then $y\ne 0$ at $x= a_\nu$.

\item  All the zeros of $y$ are {\em distinct}.

\item The zeros of $y$ lie in the interval $[a_0,a_m]$.

\end{itemize}
The case with $m=1$ corresponds to the {\em hypergeometric} differential 
equation, while the case with
$m=2$ corresponds to the {\em Heun} equation \cite{PBD}.
For $m\le 3$ polynomial solutions mostly characterize QES models
\cite{Trb1,Tcrc,Tams,SmTr,Zh,TAP,FGR}, although not all such polynomial 
solutions are exhausted by the QES models \cite{Zh}.
General (extended) Heine-Stieltjes polynomials were often 
studied in connection with a special {\em Lipkin-Meshkov-Glick} model 
corresponding to the standard two-site Bose-Hubbard model \cite{PBD,ML}.

Compared to the continuous case of Eq. (\ref{s6.8.1}), very little is known
about general properties of polynomial solutions of a linear homogeneous 
second-order finite-difference equation
\begin{equation}
g(x)\Dth^2 y(x) + r(x)\Dth y(x) + u(x)y(x+h) = 0,
\label{l9}
\end{equation}
where the first difference quotient of $y(x)$, or N\"{o}rlund's 
operator $\Dth$ \cite{Nrb,MT}, is defined here in usual sense
\begin{equation}
\Dth y(x)={y(x+h)-y(x)\over h}\cdot
\nonumber
\end{equation}
The finite-difference equation (\ref{l9}) can be disguised in further equivalent forms 
\begin{eqnarray}
g(x)\Dth^2 y(x) + [r(x)+hu(x)]\Dth y(x) + u(x) y(x) &=& 
\nonumber\\
g(x)y(x+2h) + [hr(x)+h^2 u(x)-2g(x)]y(x+h) + [g(x)-hr(x)] y(x)=0.
\label{l9f}
\end{eqnarray}
(The first one follows on making use of the identity
$ay(x+h) = ha\Dth y(x) + ay(x)$.)
Last but not the least, if $y(x)=\prod_{j=1}^n (x-x_j)$ is a polynomial solution,
then Eq. (\ref{l9f}) leads at any zero $x_k$ of $y(x)$ to a 
{\em discrete Bethe Ansatz} equation (cf. Sec. 5 of Ref. \cite{INS})
\begin{equation}
 \frac{\prod_{j=1}^n (x_k-x_j+h)}{\prod_{j=1}^n (x_k-x_j-h)} = {h r(x_k-h)-g(x_k-h)
             \over g(x_k-h)}\cdot
\label{l9ba}
\end{equation}

The motivation to study polynomial solutions of finite-difference equations has 
got a boost after it was demonstrated that physical models with a discrete 
nondegenerate spectrum can be characterized
in terms of orthogonal polynomials of a {\em discrete} variable and
their weight function \cite{Lnz,Hd,AMops,AMtb,AMef,AMhd}.
The latter applies to all problems where Hamiltonian operator 
is a self-adjoint extension of a tridiagonal Jacobi matrix of deficiency index 
$(1,1)$ \cite{FuH}. For instance a displaced harmonic oscillator can be characterized
in terms of the classical {\em Charlier} polynomials and the Rabi model
by a norm preserving deformation of the Charlier polynomials \cite{AMops,AMtb}.
Some earlier applications of classical discrete polynomials in physics not related to 
Lanczos-Haydock scheme \cite{Lnz,Hd} have been 
given by Lorente \cite{Lor}. He showed that the respective 
orthonormal {\em Kravchuk} and {\em Meixner} functions are related to a 
quantum harmonic oscillator 
and the hydrogen atom of discrete variable, and that the {\em Hahn} polynomials are related to
Calogero-Sutherland model on the lattice.

Unfortunately, only the hypergeometric case $m=1$, where the polynomial 
coefficients $g(x), r(x), u(x)$ have degrees $2, 1, 0$, respectively, 
has been studied exhaustively within the realm of 
{\em classical} orthogonal polynomials of a {\em discrete} variable
\cite{Lnc,Hn4,Lsk5,KLS}. Generalized Bochner theorem for finite-difference equations 
has been dealt with in Ref. \cite{VZ}. 
An important step forward has been achieved by Turbiner \cite{Trb1,Tcrc,Tams,SmTr}
within the realm of {\em quasi-exactly-solvable} (QES) equations
\cite{Trb1,Tcrc,Tams,Zh,SmTr,TAP}. The latter yield a specific subclass 
of finite-difference equations (\ref{l9}) where the polynomial 
coefficients $g(x), r(x), u(x)$ have degree at most {\em four}.

The motivation of present work is to translate the properties of the classical continuous 
Heine-Stieltjes theory into the realm of finite-difference equations.
As in the continuum case, a second order finite-difference equation (\ref{l9}) 
has {\em two} linearly independent solutions for a fixed triplet of polynomial 
coefficients $g(x), r(x), u(x)$. Here we derive the conditions under which 
two linearly independent {\em polynomial} solutions of Eq. (\ref{l9}) are forbidden, i.e. the 
polynomial solutions of {\em general} second-order 
finite-difference equation (\ref{l9}) are {\em unique} (cf. Theorems 1 and 2).
As a by product, an $h$-analogue of {\em Abel}'s theorem for the Heine-Stieltjes problem 
is derived, which yields an explicit
analytic expression of finite difference {\em Wronskian}, or {\em Casoratian}, $W_h(x)$
in terms of a rational function involving products of generalized gamma function 
$\Gamma_h(x)$ in both numerator and denominator.
A comparison with the classical hypergeometric equation is provided in Sec. \ref{sc:hypo}.
Using an intrinsic relation between the Heine-Stieltjes problem problem (\ref{l9}) and the 
{\em discrete} Bethe Ansatz equations (\ref{l9ba}),
the uniqueness result is extended  from the former to the latter in Sec. \ref{sc:bane}.
The results are discussed from different perspectives in Sec. \ref{sc:dsc}.
Sec. \ref{sc:rcvu} shows that the uniqueness in our sense ensures uniqueness
even if the {\em regularity condition} for the Hahn class of hypergeometric 
orthogonal polynomials (cf. Sec. 2.3 of Ref. \cite{KLS}) does not preclude
two polynomial solutions. A comparison with  one of the Shapiro problems is discussed 
in Sec. \ref{sc:shp}. 
There are two basic ways how to make use of the uniqueness theorems for 
Heine-Stieltjes polynomials. First, they yield a straightforward proof 
of the {\em nondegeneracy} of QES levels which yield
the so-called {\em exceptional spectrum} of physical models 
\cite{Trb1,Tcrc,Tams,SmTr,TAP,Hd,AMhd}, 
the proof of which is more involved by other means (cf. Refs. \cite{Hd,AMhd,FGR}). 
Second, they serve as {\em no-go} theorems in certain exceptional 
cases - cf. Sec. \ref{sc:dsc}.

\section{Uniqueness}
For each $x\in\mathbb{R}$ one can define the lattice 
$\Lambda_{h}(x):=\{x+kh \, |\, k\in \mathbb{Z}\}$.
For a given $x_0\in\mathbb{R}$ the second order finite-difference equation 
(\ref{l9f}) is seen to connect the 
values of $y(x)$ at the points of $\Lambda_{h}(x_0)$. 
The function constant on each $\Lambda_{h}(x_0)$ is called $h$-{\em periodic} function. 
Two functions $y_1$ and $y_2$ are called {\em linearly dependent} 
in a finite-difference sense if there are 
$h$-{\em periodic} functions $C_1$ and $C_2$ such that 
$C_1(x) y_1(x)+ C_2(x) y_2(x)\equiv 0$. Otherwise the 
functions $y_1$ and $y_2$ are called {\em linearly independent}.
We shall use repeatedly the following elementary argument:
If $y(x)$ is known to be a polynomial of degree not larger than $N$
and, at the same time, to vanish in at least $N+1$ different points, then $y(x)\equiv 0$.
In what follows we shall consider the second order finite-difference equation 
(\ref{l9f}) with polynomial coefficients only.
Using the argument, one finds immediately that:
\begin{itemize}

\item ({\bf P1}) If a polynomial $y(x)$ solves equation (\ref{l9f}) on an {\em infinite}
subset of $\Lambda_{h}(x_0)$, then $y(x)$ solves it for all $x_0\in\mathbb{R}$.

\item ({\bf P2}) If a linear combination $C_1 y_1(x)+ C_2 y_2(x)$ of
two polynomials vanishes on an {\em infinite}
subset of $\Lambda_{h}(x_0)$, then it vanishes for all $x\in\mathbb{R}$.

\end{itemize}
The latter implies that linear dependence of two polynomial solutions $y_1$ and $y_2$ 
in a {\em finite-difference} sense reduces to the linear dependence 
in {\em conventional} sense, i.e. with $C_1$ and $C_2$ being {\em independent} of $x$. 

The following theorem, and Theorem 2 below, encompass 
all {\em quasi-exactly-solvable equations}  
on a uniform linear-type lattice \cite{Trb1,Tcrc,Tams,SmTr}
and all classical orthogonal polynomials of a discrete variable \cite{Lnc,Hn4,Lsk5,KLS}.
\vspace*{0.4cm}

{\bf Theorem 1:}
Let the second-order finite-difference equation (\ref{l9})
has polynomial coefficients such that $g(x)$ and $g(x)- hr(x)$ have 
{\em real} roots, 
\begin{eqnarray}
g(x)  &=& (x- b_0)(x- b_1)\ldots (x- b_{m'}),\qquad b_0< b_1 < \ldots  < b_{m'},
\nonumber\\
g(x)- hr(x) &=& (x- a_0)(x- a_1)\ldots (x- a_m),\qquad a_0< a_1 < \ldots  < a_m.
\label{s6.81.1d} 
\end{eqnarray}
For each root $a_j$ define a uniform lattice 
$\Lambda_{a_j}:=\{ a_j+ kh\, |\, k\in \mathbb{N}_0\}$,
which extends to the right of the root $a_j$.
It is not excluded that $a_{j_2},\, b_l\in \Lambda_{a_{j_1}}$ for $a_{j_2},\, b_l>a_{j_1}$.
Assume further that there is at least a {\em single} $\Lambda_{a_j}$ which 
does not contain any root of $g(x)$.
Then Eq. (\ref{l9}) cannot have two polynomial
solutions $y_1$ and $y_2$ linearly {\em independent} of each other.
\\

{\em Proof}:

For any two functions $y_1$ and $y_2$ the {\em Leibniz}'s theorem 
of finite-difference calculus (pp. 34-35 of Milne-Thomson \cite{MT}) implies
\begin{equation}
\Dth [y_1(x+h)\cdot \Dth y_2(x)- \Dth y_1(x)\cdot y_2(x+h)]=
y_1(x+h)\cdot \Dth^2 y_2(x)- \Dth^2 y_1(x)\cdot y_2(x+h).
\nonumber
\end{equation}
Hence for two nontrivial solutions $y_1$ and $y_2$ of the 
finite-difference equation (\ref{l9}) we have
\begin{eqnarray}
\lefteqn{
g(x)\Dth[y_1(x+h)\cdot \Dth y_2(x)- \Dth y_1(x)\cdot y_2(x+h)] 
}
\nonumber\\
     && ~~~~~~~~   + r(x)[y_1(x+h)\cdot \Dth y_2(x)- \Dth y_1(x)\cdot y_2(x+h)] = 0.
\label{l9p}
\end{eqnarray}
The latter is of the form
\begin{equation}
\Dth X(x) = - {r(x) \over g(x)}\, X(x)~~~~\mbox{or}~~~~
X(x+h) = \frac{g(x)- hr(x)}{g(x)}\, X(x)= R(x)X(x),
\label{Xfde}
\end{equation}
where $X$ stands for the square bracket in Eq. (\ref{l9p}), which can be 
identified with a finite difference {\em Wronskian}, or {\em Casoratian}, \cite{KPe}
\begin{eqnarray}
W_h\{y_1,y_2\}(x) &:=& 
\left\vert
\begin{array}{cc}
y_1(x+h) & y_2(x+h) 
\\
\Dth y_1(x) & \Dth y_2 (x)
\end{array}
\right\vert
= {1\over h}
\left\vert
\begin{array}{cc}
y_1(x) & y_2(x) 
\\
y_1(x+h) & y_2(x+h) 
\end{array}
\right\vert
\nonumber\\
 &=&
y_1(x)\cdot \Dth y_2(x) - \Dth y_1(x)\cdot y_2(x).
\label{Prtc}
\end{eqnarray}
The hypotheses of Theorem 1 determine $R(x)$ as 
a rational function with zeros and poles on the real axis
\begin{equation}
R(x):=\frac{g(x)- hr(x)}{g(x)}={\prod_{j=0}^m (x-a_j) \over \prod_{l=0}^{m'} (x-b_l)}\cdot
\label{ff}
\end{equation}
Now if $\Lambda_{a_j}$ does not contain any zero of $g(x)$, i.e. 
$R(x)$ is not singular on $\Lambda_{a_j}$, then, in virtue of  $R(a_j)=0$,
the first-order recurrence (\ref{Xfde}) implies $X(x)\equiv 0$ for all $x\in \Lambda_{a_j+h}$.
In other words, for each $x_s \in \Lambda_{a_j+h}$ there are
$C_1(x_s)$ and $C_2(x_s)$ not both zero, such that 
$C_1(x_s) y_1(x_s)+ C_2(x_s) y_2(x_s)=C_1(x_s) y_1(x_s+h)+ C_2(x_s) y_2(x_s+h)= 0$.
Taking $x_s=a_j+h$, the linear combination $y(x):=C_1(x_s)y_1(x)+ C_2(x_s) y_2(x)$ 
is a solution of Eq. (\ref{l9})
on $\Lambda_{a_j+h}$ which satisfies $y(x_s)=y(x_s+h)=0$.
Considering the latter as the initial values of the {\em Cauchy problem}
for the recursive form (\ref{l9f}) of Eq. (\ref{l9}), one has 
$y(x)\equiv 0$ on $\Lambda_{a_j+h}$.
Because $g(x)\ne 0$, the solutions of the 
Cauchy problem for Eq. (\ref{l9f}) are {\em uniquely} determined 
by the initial values \cite{KPe}, and hence 
$C_1$ and $C_2$ are constants on entire $\Lambda_{a_j+h}$.
In virtue of the elementary argument ({\bf P2}), 
the linear combination $C_1 y_1(x)+ C_2 y_2(x)$ 
vanishes for all $x\in\mathbb{R}$, i.e. $y_1$ and $y_2$ are 
{\em linearly dependent} in the {\em conventional} sense. 
\vspace*{0.4cm}

{\bf Remark:} On considering equation (\ref{Xfde})
as a {\em downward} recurrence $X(x)=R^{-1}(x) X(x+h)$, an alternative version
of Theorem 1 follows which guarantees the uniqueness, provided that there is at least a 
{\em single} $\Lambda_{b_l}$ which does not contain any root of $g(x)- hr(x)$.
Here $\Lambda_{b_l}$ is defined for each root $b_l$ as a uniform lattice 
which extends to the {\em left} of the root $b_l$,
$\Lambda_{b_l}:=\{ b_l- kh\, |\, k\in \mathbb{N}_0\}$.

\vspace*{0.4cm}

{\bf Theorem 2:}
Let us consider the second-order finite-difference equation (\ref{l9})
with the polynomial coefficients as in Theorem 1.
Assume further that there is at least a {\em single} $\Lambda_{a_j}$ which contains 
more roots (e.g. the single root $a_j$) of $g(x)- hr(x)$ than the roots of $g(x)$ 
[e.g. none of the roots $b_l$ of $g(x)$].
Then Eq. (\ref{l9}) cannot have two polynomial
solutions $y_1$ and $y_2$ linearly {\em independent} of each other.
\vspace*{0.4cm}

Before giving the proof of Theorem 2, it is expedient 
to provide an $h$-analogue of {\em Abel}'s 
theorem which yields an explicit
analytic expression of $X(x)$ in terms of a rational function involving 
products of $\Gamma_h$ in both its numerator and denominator. 
The $h$-extension of the gamma function $\Gamma_h(x)$  
is introduced through the functional equation $\Gamma_h(x+h)= x \Gamma_h(x)$ 
(cf. sec. 9.66 of Ref. \cite{MT}; Appendix \ref{sc:ggmf}).
\vspace*{0.4cm}

{\bf Lemma 1:}
For any rational $R(x)$ of the form (\ref{ff}), 
the solution $X(x)$ of the first-order finite-difference equation 
(\ref{Xfde}) is either identically zero or
\begin{equation}
X(x) = \mbox{const}\times  
   \frac{\prod_{j=0}^m\Gamma_h(x-a_j) }{\prod_{l=0}^{m'} \Gamma_h(x-b_l)}\cdot
\label{l9psfe}
\end{equation}
Provided that the ratio $\kappa$ of the leading polynomial coefficient of
$g(x)- hr(x)$ to that of $g(x)$ is $\kappa\ne 1$, the r.h.s. of Eq. (\ref{l9psfe})
will acquire an additional multiplication factor [cf. Eq. (\ref{nbi1})]
and becomes
\begin{equation}
X(x) = \mbox{const}\times \kappa^{x-(h/2)} \frac{\prod_{j=0}^m\Gamma_h(x-a_j) }
                 {\prod_{l=0}^{m'} \Gamma_h(x-b_l)}\cdot
\label{l9psfef}
\end{equation}
\\

{\em Proof}:

First, Eq. (\ref{Xfde}) is recast as
\begin{equation}
\Dth \ln X  ={1 \over h} \ln \left( \frac{g(x)- hr(x)}{g(x)} \right),
\nonumber
\end{equation}
which has the form of the first-order 
finite-difference equation (\ref{fd1}). Its solution can be expressed in terms of 
N\"{o}rlund's {\em principal solution} \cite{Nrb,MT}, an elegant, 
but nowadays largely forgotten, 
tool of integrating finite-difference equations
(see Appendix \ref{sc:npcs} for a brief summary and definition), as 
\begin{equation}
X(x) = \exp\left[\sint_0^x {1\over h}\, 
       \ln \left( \frac{g(t)- hr(t)}{g(t)} \right) \Dth t\right].
\label{l9ps}
\end{equation}

Note in passing that use of a partial fraction decomposition (\ref{s6.81.2}) of the 
fraction in the integrand in the exponent of Eq. (\ref{l9ps}), as in the continuous case 
of Stieltjes \cite{Stl} and further elaborated in Sec. 6.81 of Ref. \cite{Sz}, 
would not bring us any further. Instead it is expedient to 
substitute the respective products (\ref{s6.81.1d}) into Eq. (\ref{l9ps}) and use
the logarithm there to split the resulting ratio into a sum of
individual logarithms $\ln (t-a_j)$ and $-\ln (t-b_l)$ corresponding to the
roots in Eq. (\ref{s6.81.1d}). Each such a logarithm term
integrates to a corresponding {\em generalized gamma function} 
$\Gamma_h$ (cf. Eq. (\ref{gfdf2}) of Appendix \ref{sc:ggmf};
sec. 9.66 of Ref. \cite{MT}). The latter recipe 
enables one to express (\ref{l9ps}) as in Eq. (\ref{l9psfef}).
The transition from (\ref{l9ps}) to (\ref{l9psfef}) is 
similar to that used by Lancaster \cite{Lnc} in arriving from 
his Eq. (29) to his Eqs. (30-33).
\vspace*{0.4cm}

{\em Proof of Theorem 2}:

If $X(x)$ of two linearly independent solutions in Eq. (\ref{Xfde}) 
is not identically zero, the hypotheses of Theorem 2 imply that
$X$ is necessarily {\em singular} for 
some its argument value, which is impossible if $y_1$ and $y_2$ are polynomials.
Indeed, the hypotheses of Theorem 2 ensure that there is at least a {\em single} 
$\Lambda_{a_j}$ which contains more roots (e. g. the single root $a_j$) of $g(x)- hr(x)$ 
than the roots of $g(x)$ (e. g. none of the roots $b_l$ of $g(x)$). 
Unless $X(x)$ is identically zero, Lemma 1 determines
the analytic form of $X(x)$ to be either (\ref{l9psfe}) or (\ref{l9psfef}).
Now $\Gamma_h(x-a_j)$ has a simple pole at $x=a_j$ (cf. Appendix \ref{sc:ggmf}). 
If there is $a_j<a_k\in\Lambda_{a_j}$, then also $\Gamma_h(x-a_k)$ has a simple pole 
at $x=a_j$. If there is $b_l\in \Lambda_{a_j}$, some of 
the simple poles of $\Gamma_h(x-a_j)$ and $\Gamma_h(x-a_k)$ at 
$x=a_j$ in the numerator on the r.h.s. of Eq. (\ref{l9psfef}) could be canceled 
by the simple pole of the $\Gamma_h(x-b_l)$ at $x=a_j$ 
in the denominator on the r.h.s. of Eq. (\ref{l9psfef}).
Nevertheless, the hypotheses of Theorem 2 guarantee that at least one of the simple poles 
of $\Gamma_h$'s in the numerator is not compensated by the simple pole 
of $\Gamma_h(x-b_l)$ in the denominator. Then $X(x)$ tends to infinity for $x\to a_j$.
However, as a discrete Wronskian of two {\em polynomial} solutions, $X(x)$ 
cannot tend to infinity at any finite $x\in\mathbb{R}$. Of course, the latter does not
hold for general nonpolynomial solutions. Thus, as in the continuum 
case of Sec. 6.81 of Ref. \cite{Sz}, we have a {\em contradiction}, 
unless, of course, $X\equiv 0$.

\subsection{Classical hypergeometric equation}
\label{sc:hypo}
As an example, consider the classical hypergeometric equation \cite{Lnc,Hn4,Lsk5,KLS}
\begin{equation}
(ax^2+bx+c)\Dth^2 y(x) + (dx+f)\Dth y(x) + \lambda y(x+h)=0.
\label{fd2o}
\end{equation}
A {\em necessary} and {\em sufficient} condition for the existence 
of a polynomial solution of Eq. (\ref{fd2o}) is that a characteristic polynomial, 
\begin{equation}
\theta(z) := az(z-1)+d z +\lambda,
\nonumber 
\end{equation}
has a non-negative {\em integer} root (cf. the $n=2$ case of Theorem 2 of Ref. \cite{Lnc}).
If there is a polynomial solution of degree $n$,
then $\theta(n) =0$. The latter is equivalent to
\begin{equation}
\lambda +nd+n(n-1)a=0,~~\mbox{or}~~\lambda_n= -n(n-1)a - nd,\qquad n=0,1,2,\ldots.
\label{d20ev}
\end{equation}
Eq. (\ref{fd2o}) is a special case of the eigenvalue problems
for the {\em Hahn class} of orthogonal polynomials \cite{Hn4,KLS}. 
In the latter case the {\em regularity condition} says that all 
eigenspaces of the hypergeometric eigenvalue problem are one dimensional 
if and only if $\lambda_n\ne \lambda_l$ for $l\ne n$ 
in the set of numbers $\{\lambda_n\}_{n=0}^\infty$ defined by Eq. (\ref{fd2o}), or 
if and only if $a[n]+d \ne 0$ (cf. Sec. 2.3 of Ref. \cite{KLS}).
Here $[-1]=-1/q$, $[0]=0$, $[n]=\textstyle\sum_{k=0}^{n-1} q^k$, $n\ge 1$, and 
$q\in \mathbb{R}\backslash \{-1,0\}$ is the Hahn parameter (for the uniform linear lattice
in our case $q=1$ and $[n]=n$).
However, the {\em regularity condition} does not exclude 
the corresponding eigenspace to be, for instance, two dimensional 
for $\lambda_n=\lambda_l$ with $l\ne n$. The latter is precluded by the following
Corollary.
\vspace*{0.4cm}

{\bf Corollary:} Polynomial solutions of the second-order 
finite-difference hypergeometric equation Eq. (\ref{fd2o}) are {\em nondegenerate}, i.e.,
for a given eigenvalue $\lambda$ there is at most a single solution to Eq. (\ref{fd2o}).
\\

If $d=-ka$, then $\lambda_n=-n(n-k-1)a$ and $\lambda_n$ may equal $\lambda_l$ 
for some $l\ne n$. For instance, $\lambda_1=\lambda_k=k$ for $n=1,k$. 
If $k>1$ there is thus, under the hypotheses of Theorems 1 and 2, 
no polynomial solution of degree $k$.

\subsection{Discrete Bethe Ansatz equations}
\label{sc:bane}
Using an intrinsic relation between the Heine-Stieltjes theory and the 
{\em discrete} Bethe Ansatz equations one can immediately arrive at the
following result.
\vspace*{0.4cm}

{\bf Theorem 3:}
Provided that the Heine-Stieltjes problem has unique polynomial
solution, the corresponding {\em discrete} Bethe Ansatz equations 
(\ref{l9ba}) have also a unique polynomial solution up to permutations
of zeros $x_k$'s.
\\

{\em Proof}:

A solution $y(x)=\sum_{j=0}^n y_jx^j$ to the {\em discrete} 
Bethe Ansatz equations (\ref{l9ba}) 
implies that the second-order difference equation (\ref{l9}) 
is satisfied at the $n$ points $x_1,x_2,\ldots, x_n$.
The necessary condition that $y(x)=\sum_{j=0}^n y_jx^j$ solves Eq. (\ref{l9}) 
is the vanishing of the leading $n$th degree. 
The latter requires that the sum of the coefficients of the leading degree of 
the polynomials 
$g(x),\, [hr(x)+h^2 u(x)-2g(x)]$, and $[g(x)-hr(x)]$ in 
the recurrence form (\ref{l9f}) of Eq. (\ref{l9}) vanishes.
In the hypergeometric case this is the condition (\ref{d20ev}).
If the polynomial coefficients 
of Eq. (\ref{l9}) are assumed 
to satisfy the necessary condition, the l.h.s. of Eq. (\ref{l9}) becomes a 
polynomial of one less, i.e. $(n-1)$th, degree.
By the elementary argument (e.g. leading to {\bf P1}),
if a polynomial in $x$ of degree $n-1$, that can vanish only 
at $n-1$ different points, vanishes at the $n$ distinct points 
$x_1,x_2,\ldots, x_n$, then it must vanish identically.
Thus the l.h.s. of Eq. (\ref{l9}) vanishes identically.
This leads to a second-order difference equation whose
polynomial solutions are unique.
\\

Ismail et al have earlier shown that the solution to the {\em discrete}
Bethe Ansatz equations (\ref{l9ba}) with the right-hand side 
derived from the Meixner $M_n(x;\beta,c)$ and 
the Hahn polynomials $Q_n(x;\alpha,\beta,N)$ are unique up to permutations
(cf. Sec. 5 of Ref. \cite{INS}). Theorem 3 extends
the results of Ismail et al (cf. Sec. 5 of Ref. \cite{INS}) to the general case.
Some special cases when the uniqueness may break down are discussed in Sec. \ref{sc:shp}.

\section{Discussion}
\label{sc:dsc}
Our uniqueness theorems encompass all {\em quasi-exactly-solvable equations}  
on a uniform linear-type lattice \cite{Trb1,Tcrc,Tams,SmTr}
 and all classical orthogonal polynomials of a discrete variable \cite{Lnc,Hn4,Lsk5,KLS}. 
The hypotheses of our uniqueness theorems look rather different 
from those in the classical  continuous 
Heine-Stieltjes theory \cite{Heine1878,Stl,Vlk,Bch,Sz}.
In the finite-difference case, the respective $g(x)$ and $g(x)- hr(x)$ 
can be identified as the coefficients of 
$y(x+2h)$ and $y(x)$ in the recurrence form (\ref{l9f}) of  Eq. (\ref{l9}). Unlike 
the continuous case of Refs. \cite{Stl,Sz} (i.e. with $\Dth$ in Eq. (\ref{l9})
replaced with ordinary derivatives [cf. Eq. (\ref{s6.8.1})]), one does not assume that 
the zeros of $g(x)$ {\em alternate} with those of $r(x)$ 
[cf. Eqs. (\ref{s6.81.1}), (\ref{s6.81.2})].
The hypotheses of Theorems 1 and 2 are also silent about relative degrees of 
the polynomial coefficients $g(x), r(x), u(x)$.

However, the above differences are mostly only apparent, until one realizes that already in the 
classical continuous Heine-Stieltjes theory the assumptions that {\bf (i)} the zeros 
of $A(x)$ {\em alternate} with those of $B(x)$ and that {\bf (ii)} the leading order coefficients 
of $A(x)$ and $B(x)$ have the {\em same} sign, are not necessary for the uniqueness of solutions.
Indeed, one can multiply both the numerator and denominator in $B(x)/A(x)$
on the l.h.s. of Eq. (\ref{s6.81.2}) with the same polynomial factor
$(x-\gamma_l)^{n_l}$, $n_l\ge 1$, without changing the r.h.s. of Eq. (\ref{s6.81.2}), 
and hence the reasoning leading to the uniqueness. 
It is also not necessary that all $\rho_\nu>0$ as in Eq. (\ref{s6.81.2}).
(The latter has been recognized as late as 2000 by Dimitrov and Van Assche \cite{DA}.)

A broader sufficient condition for the Wronskian $W\{y_1,y_2\}$ to diverge to infinity is
that there is merely at least one $\nu$ such that 
$\rho_\nu>0$ and $a_\nu$ is different from all other $b_\mu$'s.
The latter points could be illustrated for a {\em continuous} hypergeometric 
analogue of Eq. (\ref{fd2o}),
\begin{equation}
(ax^2+bx+c) y''(x) + (dx+f) y'(x) + \lambda y(x)=0.
\label{fd2oh}
\end{equation}

\subsection{Regularity condition vs uniqueness}
\label{sc:rcvu}
The regularity condition of the eigenvalue problems
for the {\em Hahn class} of orthogonal polynomials does not answer what happen
if $\lambda_n= \lambda_l$ for $l\ne n$ in the set of numbers $\{\lambda_n\}_{n=0}^\infty$ 
defined by Eq. (\ref{fd2o}). Will the eigenspace corresponding to
$\lambda_n= \lambda_l$ be zero-, one-, or two-dimensional?
The question of uniqueness and existence of the polynomial solutions of the 
{\em hypergeometric} equation (\ref{fd2oh}) reduces to solving 
Lesky's {\em downward} TTRR (cf. Eq. (3) in Ref. \cite{Lsk})
\begin{eqnarray}
(n-k)[(n+k-1)a+d] a_{nk} = (k+1)[(k+2)c\, a_{n,k+2} + (kb+f) a_{n,k+1}]
\label{lttr}
\end{eqnarray}
for the coefficients $a_{nk}$ of the polynomial solution of the $n$th degree,
\begin{equation}
y(x)=a_{nn} x^n + a_{n,n-1}x^{n-1} +\ldots + a_{n0}.
\end{equation}
The TTRR runs {\em downward} for $k=n-1,n-2,\ldots, 0$, with the initial condition 
$a_{n,n+1}\equiv 0$. Without any loss of generality one can assume $a_{nn}=1$. 
With the initial conditions on $a_{n,n+1}$ and $a_{nn}$ being fixed, any 
other not linearly dependent solution 
has to have $a_{n,n+1}\ne 0$ for $W_h(x)$ [see Eq. (\ref{Prtc})] of the Cauchy problem 
for the TTRR (\ref{lttr}) to be nonzero.
This is impossible for a polynomial solution of the $n$th degree,
which implies uniqueness of the polynomial solution of the $n$th degree, provided
it exists (i.e. $(n+k-1)a+d\ne 0$).

The condition (\ref{d20ev}) is valid both in the continuous and discrete cases.
Thus for $d=-ka$ some of $\lambda_n$ may equal $\lambda_l$ also in the continuous case
(e.g. $\lambda_1=\lambda_k=k$ for $n=1,k$). Let $\lceil x \rceil$ denotes the smallest 
integer not less than $x$, or the {\em ceiling} function.
Then, unless some additional conditions are satisfied, Lesky's TTRR (\ref{lttr}) does not have 
any solution for $d=-ka$ and $n\in [\lceil (k+2)/2 \rceil,k+1]$. 
Obviously the uniqueness of polynomial solutions persists even though the above 
assumption {\bf (ii)} is not satisfied.

\subsection{Shapiro problem}
\label{sc:shp}
The additional conditions under which Lesky's TTRR (\ref{lttr}) has a solution 
for any degree $n$ and when the uniqueness of polynomial solutions breaks down 
are formulated separately for $k$ even and odd.
The latter is related to the problem of describing when a linear ordinary 
differential equation with polynomial coefficients admits 
at least $2$ polynomial solutions, which is the first of five open problems 
listed by Shapiro \cite{Shp}. An exhaustive answer 
in the special case $2B(x)=-A'(x)$ has been obtained by Eremenko and Gabrielov \cite{EG}.
The following discussion is limited to the 
{\em hypergeometric} equation (\ref{fd2oh}) but is not constraint to $2B(x)=-A'(x)$.

For $d=-ka$ and even $k=2t>0$ (Lesky's special case 2), uniqueness persists unless $f=-tb$. 
Then $dx+f=-2tax-tb$, or $2B(x):=dx+f=-tA'(x)$ in the notation of Eq. (\ref{s6.8.1}), and hence
all the residues $\rho_j$ of the ratio $B(x)/A(x)=-tA'(x)/[2A(x)]$ 
in Eq. (\ref{s6.81.2}) are necessarily {\em negative}.
For odd $k=2t-1>0$ (Lesky's special case 3), the ratio $B(x)/A(x)=-t/(x-a_0)-(t-1)/(x-a_1)$,
i.e. none of the residues $\rho_j$ of the ratio $B(x)/A(x)$ is positive.
Thus not just any {\em algebraic dependence} of $A(x)$ and $B(x)$ but only a particular one
\cite{Shp} leads to that the uniqueness of polynomial solutions ceases to hold and there 
are possible two linearly independent solutions of the continuous 
Eq. (\ref{fd2oh}) for the same value of $\lambda$.

\section{Conclusions}
\label{sc:cncl}
We have established sufficient conditions (Theorems 1 and 2) for the uniqueness 
of polynomial solutions of second order finite-difference equations. 
They encompass 
all classical orthogonal polynomials of a discrete variable
\cite{Lnc,Hn4,Lsk5,KLS} and all {\em quasi-exactly-solvable equations} on a uniform
linear-type lattice \cite{Tcrc,Tams,SmTr}.
An $h$-analogue of {\em Abel}'s theorem for the Heine-Stieltjes problem 
was derived, which yields an explicit
analytic expression of finite difference {\em Wronskian}, or {\em Casoratian}, $W_h(x)$
in terms of a rational function involving products of generalized gamma function 
$\Gamma_h(x)$ in both numerator 
and denominator. The latter was facilitated
by N\"{o}rlund's {\em principal solution}~ $\sint$ \cite{Nrb,MT}.
It suffices to know N\"{o}rlund's principal solution~ $\sint$
only for a constant [cf. Eq. (\ref{nbi1})] and a logarithm [cf. Eq. (\ref{gfdf2})]
to deal with a large set of finite difference problems (e.g. Ref. \cite{Lnc}).
Using an intrinsic relation between the Heine-Stieltjes problem (\ref{l9}) and the 
{\em discrete} Bethe Ansatz equations (\ref{l9ba}), Theorem 3 extended
the uniqueness of polynomial solutions of the discrete Bethe Ansatz equations 
of Ismail et al (cf. Sec. 5 of Ref. \cite{INS}) 
to the general case.
The uniqueness in the classical continuous Heine-Stieltjes theory was shown to 
hold under broader hypotheses than usually presented \cite{Stl,Bch,Sz}. 
A difference between the regularity condition and uniqueness was emphasized.
An extension of the results to a general lattice and a second-order 
finite-difference equation (\ref{l9})
with $\Dth$ being replaced by the more general Hahn operator \cite{Hn4,KLS}
is dealt with in a forthcoming publication \cite{AMhs}.
An open question remains if it is possible to translate
also the remaining properties of the classical continuous Heine-Stieltjes theory 
into the realm of finite-difference equations.

\section{Acknowledgment}
Continuous support of MAKM is greatly acknowledged.
I am also indebted to Y.-Z. Zhang for providing me with Lancaster work \cite{Lnc}
and A. V. Turbiner for providing me with copies of some of his work and for a discussion.

\newpage 

\appendix

\section{N\"{o}rlund's  principal solution}
\label{sc:npcs}
That particular solutions of the given equation 
\begin{equation}
\Dth u(x)=\phi(x),
\label{fd1}
\end{equation}
always exist is seen (in the case of the real variable) by considering that $u(x)$ being
arbitrarily defined at every point of the interval $0\le x<h$, the
equation defines $u(x)$ for every point exterior to this interval. 
The expression
\begin{eqnarray}
f(x) &=& A-h [\phi(x) + \phi(x+h)+ \phi(x+2h)+ \phi(x+3h)+\ldots]
\nonumber\\
&=& A-h \sum_{s=0}^\infty \phi(x+sh),
\nonumber
\end{eqnarray}
where $A$ is constant, is a formal solution of the difference equation, since
\begin{equation}
f(x+h) = A-h [\phi(x+h)+ \phi(x+2h)+ \phi(x+3h)+\ldots],
\nonumber
\end{equation}
and therefore  $f(x+h)-f(x) = h\phi(x)$.
However, such solutions are in general {\em not} analytic. 

N\"{o}rlund \cite{Nrb} has succeeded in defining
a {\em principal solution} which has specially simple and definite properties. 
In particular, when $\phi(x)$ is a polynomial so is the principal
solution. If for $A$ we write $\int_c^\infty \phi(t) dt$, and if this infinite integral and the
infinite series both converge, N\"{o}rlund defines the {\em principal solution} of the
difference equation, or sum of the function $\phi(x)$, as
\begin{equation}
F(x) = \sint_c^x \phi(z)\, \Dth z = \int_c^\infty \phi(t)\, dt -h \sum_{s=0}^\infty \phi(x+sh).
\label{psd}
\end{equation}
The principal solution thus defined depends on an arbitrary constant $c$. 
As an example, consider \cite{Nrb,MT}
\begin{equation}
\Dth u(x)=e^{-x},
\nonumber
\end{equation}
$x$ and $h$ being real and positive. Here
\begin{eqnarray}
F(x) &=& \sint_c^x e^{-z}\, \Dth z = \int_c^\infty e^{-t}\, dt -h \sum_{s=0}^\infty e^{-x-sh}
\nonumber\\
&=& e^{-c} - {he^{-x} \over 1-e^{-h}},
\label{ppsd}
\end{eqnarray}
after evaluating the integral, and summing the geometrical progression.

The necessary and sufficient conditions for the existence of the
sum $F(x)$ as defined above are the convergence of the integral
and of the series. In general, {\em neither} of these conditions is satisfied and the definition
fails. In order to extend the definition of the sum, N\"{o}rlund adopts
an ingenious and powerful recipe. This consists in a {\em regularization} of
$\phi(x)$ with a parameter $\mu$ ($> 0$), say $\phi(x,\mu)$,
which is so chosen that 
(see Chapter III of Ref. \cite{Nrb}; see also Chapter VIII of Ref. \cite{MT})

\begin{itemize}

\item (i) $\lim_{\mu\rightarrow 0} \phi(x,\mu)=\phi(x)$;

\item (ii) $\int_c^\infty \phi(t)\, dt$ and  $\sum_{s=0}^\infty \phi(x+sh)$ both converge.

\end{itemize}
For this function $\phi(x,\mu)$, the difference equation
\begin{equation}
\Dth u(x)=\phi(x,\mu),
\label{fd5}
\end{equation}
has a principal solution, given by the definition (\ref{psd}),
\begin{equation}
F(x,\mu) =  \int_c^\infty \phi(t,\mu)\, dt -h \sum_{s=0}^\infty \phi(x+sh,\mu).
\nonumber 
\end{equation}
If in this relation we let $\mu\rightarrow 0$, the difference equation (\ref{fd5})
becomes the difference equation (\ref{fd1}) and the principal solution of
the latter is defined by
\begin{equation}
F(x) = \lim_{\mu\rightarrow 0} F(x,\mu),
\nonumber 
\end{equation}
provided that this limit exists uniformly and, subject to conditions
(i) and (ii), is {\em independent} of the particular choice of $\phi(x,\mu)$.
When the limit exists $\phi(x)$ is said to be {\em summable}.

The success of the method of definition just described
depends on the difference of the infinite integral and the infinite
series having a limit when $\mu\rightarrow 0$. Each separately may diverge
when $\mu=0$ and the choice of $\phi(x,\mu)$ has to be so made that when
we take the difference of the integral and the series the divergent
part disappears. It has been shown that, for a
wide class of summation methods, the result is {\em independent} of the
method adopted. A convenient practical choice is \cite{Nrb,MT}
\begin{eqnarray}
F(x) &=& \sint_c^x \phi(z)\, \Dth z 
\nonumber\\
&=& 
\lim_{\mu\rightarrow 0}\left\{
 \int_c^\infty \phi(t) e^{-\mu\lambda(t)}\, dt 
        -h \sum_{s=0}^\infty \phi(x+sh) e^{-\mu\lambda(x+sh)} \right\},
\label{psdrs}
\end{eqnarray}
where $p\ge 1,\, q \ge 0$, such that for $\lambda(x) = x^p(\ln x)^q$
this limit exists.
N\"{o}rlund's recipe (\ref{psdrs})  can be seen as a two-parameter extension 
of the single-parameter Lindel\"{o}f and Mittag-Leffler methods of summing divergent series \cite{H}.
The latter belongs to the so-called {\em analytic} and {\em regular} summability 
methods \cite{H,AMcjm}. If applied to a power series 
(i) it yields the value equal to that obtained by an 
analytic continuation of the series beyond the radius of convergence anytime the limit exists,
(ii) provided that the sum converges for $\mu=0$, the limit $\mu\to 0$ 
yields the very same sum \cite{H,AMcjm}.

As a simple illustration, consider
\begin{equation}
\Dth u(x)=a,
\nonumber 
\end{equation}
where $a$ is constant.
The series $a+a+a+\ldots$ obviously diverges, but for $\mu> 0$
\begin{equation}
\int_c^\infty a e^{-\mu t}\, dt,\qquad \sum_{s=0}^\infty a e^{-\mu(x+sh)}
\nonumber 
\end{equation}
both converge if $h$ is a positive real number, so that we can take $\lambda(x)=x$,
i.e. $p = 1$, $q = 0$. Hence
\begin{eqnarray} 
\sint_c^x a \, \Dth z  &=& \lim_{\mu\rightarrow 0} \left\{  
\int_c^\infty a e^{-\mu t}\, dt-h \sum_{s=0}^\infty a e^{-\mu(x+sh)} \right\}
\nonumber\\
&=& \lim_{\mu\rightarrow 0} \left(
{a e^{-\mu c} \over\mu}-{ah e^{-\mu x} \over 1-e^{-\mu h}} \right)
\nonumber\\
&=& \lim_{\mu\rightarrow 0} a e^{-\mu c} \left[
{ 1-e^{-\mu h} -\mu h e^{-\mu (x-c)} \over \mu(1-e^{-\mu h}) } \right]
\nonumber\\
&=& \lim_{\mu\rightarrow 0} 
{
a e^{-\mu c}\left[ \mu h  -{(\mu h)^2 \over 2} + \ldots 
- \mu h+\mu^2 h(x-c)- \ldots \right]
\over 
\mu\left[\mu h  -{(\mu h)^2 \over 2} + \ldots \right]
}
\nonumber\\
&=& 
a\left(x-c-{h\over 2}\right),
\label{nbi1}
\end{eqnarray} 
which is the {\em principal} solution. 
It should be noted that both the
integral and the series diverge when $\mu=0$.

\section{The generalized Gamma function}
\label{sc:ggmf}
Following sec. 9.66 of Ref. \cite{MT},
if we define the function $\Gamma_h(x)$ by the relation
\begin{equation}
h \ln \Gamma_h(x) = \sint_0^x \ln z \Dth z +h\ln\sqrt{2\pi/h},
\label{gfdf}
\end{equation}
we have by differencing
\begin{equation}
h \Dth \ln \Gamma_h(x) :=  \ln  {\Gamma_h(x+h)\over \Gamma_h(x)}=\ln x,
\label{gfdf2}
\end{equation}
and hence
\begin{equation}
 \Gamma_h(x+h)= x \Gamma_h(x).
\end{equation}
Thus, if $n$ be a positive integer, $\Gamma_h(nh+h) = h^n n! \Gamma_h(h)$.
$\Gamma_h(x)$ can be related to the conventional
$\Gamma(x)$ through
\begin{equation}
\ln \Gamma_h(x) = \ln  \Gamma(x/h) +{1\over h}\, (x-h)\ln h,
\nonumber 
\end{equation}
or
\begin{equation}
\Gamma_h(x) = \Gamma(x/h)\exp\left({x-h\over h}\,\ln h\right).
\nonumber 
\end{equation}
Using the above relation one finds $\Gamma_h(h) = 1$,
and for any positive integer $n>0$
\begin{equation}
 \Gamma_h(nh+h) = h^n n!.
\nonumber 
\end{equation}
The formula (sec. 9.66 of Ref. \cite{MT})
\begin{equation}
{1\over  \Gamma_h(x)} = e^{ {\gamma-\ln h\over h}\, x} 
   x \prod_{s=1} \left({x\over sh} +1\right) e^{ - {x\over sh} },
\end{equation}
where $\gamma\approx 0.5772$ is the Euler-Mascheroni constant, shows that $1/\Gamma_h(x)$ is 
an integral transcendent function,
with simple {\em zeros} at the points $0,-h,-2h,-3h,\ldots$, and 
therefore that $\Gamma_h(x)$ is a {\em meromorphic} function of $x$ with simple {\em poles}
at the same points.

\newpage 



\begin{thebibliography}{99}

\bibitem{Heine1878}
Heine E 1878 
{\it Handbuch der Kugelfunktionen} vol {\bf 1} (Berlin: D. Reimer Verlag) pp 472-479 
(A translation of this often quoted passage can be found in Ref. \cite{Shp}.)


\bibitem{Stl}
Stieltjes T 1885
Sur certains polyn\^omes qui v\'erifient une equation 
diff\'erentielle lin\'eaire du second
ordre et sur la th\'eorie des fonctions de Lam\'e 
{\it Acta Math.} {\bf 8} 321-6


\bibitem{Bch}
B\^{o}cher M 1897
The roots of polynomials that satisfy certain 
differential equation of the second order
{\it Bull. Amer. Math. Soc.} {\bf 4} 256-8


\bibitem{Vlk}
van Vleck E B 1898 On the polynomials of Stieltjes 
{\it Bull. Amer. Math. Soc.} {\bf 4} 426-38 


\bibitem{Sz}
Szeg\"{o} G 1939
{\it Orthogonal Polynomials}
4th edn Amer. Math. Soc. Colloquium Publications vol 23
(Providence, RI: Amer. Math. Soc.)


\bibitem{Shp}
Shapiro B 2011
Algebro-geometric aspects of Heine-Stieltjes theory
{\it J. London Math. Soc.} {\bf 83} 36-56 
(arXiv:0812.4193)


\bibitem{MMM}
Marcell\'an F, Mart\'{\i}nez-Finkelshtein A andMart\'{\i}nez-Gonz\'alez P 2007
Electrostatic models for zeros of polynomials: Old, new, and some open problems
{\it J. Comput. Appl. Math.} {\bf 207} 258-72
(arXiv:math/0512293)


\bibitem{PBD}
Pan F, Bao L, Zhai L, Cui X and Draayer J P 2011
The extended Heine-Stieltjes polynomials 
associated with a special LMG model
{\it J. Phys. A: Math. Theor.} {\bf 44}, 395305
(arXiv:1101.1039)


\bibitem{ML}
Marquette I and Links J 2012
Generalised Heine-Stieltjes and Van Vleck polynomials 
associated with degenerate, integrable BCS models
{\it J. Stat. Mech.}  P08019 
(arXiv:1206.2988)


\bibitem{Trb1}
Turbiner A V 1992
On polynomial solutions of differential equations
{\it J. Math. Phys.} {\bf 33} 3989-93


\bibitem{Tcrc}
Turbiner A V 1995
Quasi-exactly-solvable differential equations
{\it CRC Handbook of Lie Group Analysis of Differential Equations},
vol 3 (New Trends), Chapter 12, ed N Ibragimov (CRC Press) pp 331-66 
(hep-th/9409068)


\bibitem{Tams}
Turbiner A V 1994
Lie algebras and linear operators with invariant subspace
{\it Lie Algebras, Cohomologies and New Findings in
Quantum Mechanics}, {\it Contemporary Mathematics}, vol 160, 
eds N Kamran and P Olver (Providence, RI: Amer. Math. Soc.) pp 263-310
(funct-an/9301001)


\bibitem{SmTr}
Smirnov Y and Turbiner A V 1995
Lie-algebraic discretization of differential equations
{\it Mod. Phys. Lett.} A{\bf 10} 1795-802
(arXiv:funct-an/9501001)


\bibitem{Zh}
Zhang Y-Z 2012
Exact polynomial solutions of second 
order differential equations and their applications
{\it J. Phys. A: Math. Theor.} {\bf 45} 065206
(arXiv:1107.5090)


\bibitem{TAP}
Tomka M, El Araby O, Pletyukhov M, and Gritsev V
2014 Exceptional and regular spectra of the generalized Rabi model
{\it Phys. Rev. A} {\bf 90} 063839 
(arXiv:1307.7876)


\bibitem{FGR}
Finkel F, Gonz\'{a}lez-L\'{o}pez A and Rodr\'{\i}guez M A 1996
Quasi-exactly solvable potentials on the line 
and orthogonal polynomials
{\it J. Math. Phys.} {\bf 37} 3954-72
(arXiv:hep-th/9603103)


\bibitem{Nrb}
N\"{o}rlund N E 1954
{\it Vorlesungen \"{u}ber Differenzenrechnung}
(New York: Chelsea)


\bibitem{MT}
Milne-Thomson L M 1933 {\it The Calculus of Finite Differences}
(London: MacMillan)


\bibitem{INS}
Ismail M E H, I. Nikolova I and Simeonov P 2005
Difference Equations and Discriminants for Discrete Orthogonal Polynomials
{\it Ramanujan J.} {\bf 8} 475-502


\bibitem{Lnz}
Lanczos C 1950
An iteration method for the solution of the eigenvalue problem of linear 
differential and integral operators
{\it J. Res. Natl. Bur. Stand.} {\bf 45} 255-82


\bibitem{Hd}
Haydock R 1980 
The recursive solution of the Schr\"{o}dinger 
equation (Solid State Physics vol 35) ed H Ehrenreich, F Seitz and D Turnbull 
(New York: Academic) pp 215-94


\bibitem{AMops}
Moroz A 2013
On solvability and integrability of the Rabi model
{\it Ann. Phys. NY} {\bf 338} 319-40 
(arXiv:1302.2565)


\bibitem{AMtb}
Moroz A 2014
A hidden analytic structure of the Rabi model
{\it Ann. Phys. NY} {\bf 340} 252-66 
(arXiv:1305.2595) 


\bibitem{AMef}
Moroz A 2014 Quantum models with spectrum generated by the flows of polynomial zeros
{\it J. Phys. A: Math. Theor.} {\bf 47} 495204
(arXiv:1403.3773)


\bibitem{AMhd}
Moroz A 2014
Haydock's recursive solution of self-adjoint problems. Discrete spectrum
{\it Ann. Phys. NY} {\bf 351} 960-74 


\bibitem{FuH}
Fu L and Hochstadt H 1974
Inverse theorems for Jacobi matrices
{\it J. Math. Anal. Appl.} {\bf 47} 162-8


\bibitem{Lor}
Lorente M 2003
Integrable systems on the lattice and orthogonal 
polynomials of discrete variable
{\it J. Comput. Appl. Math.} {\bf 153} 321-330 
(arXiv:math-ph/0401051)


\bibitem{Lnc}
Lancaster O E 1941
Orthogonal polynomials defined by difference equations
{\it Amer. J. Math.} {\bf 63} 185-207


\bibitem{Hn4}
Hahn W 1949 
\"{U}ber Orthogonalpolynome, die $q$-Differenzengleichungen gen\"{u}gen
{\it Mathematische Nachrichten} {\bf 2} 4-34


\bibitem{Lsk5}
Lesky P 1962
\"{U}ber Polynomsysteme, die Sturm-Liouvilleschen 
Differenzengleichungen gen\"{u}gen
{\it Math. Zeit.} {\bf 78} 439-45


\bibitem{KLS} 
Koekoek R,  Lesky P A, and Swarttouw R F 
2010 {\it Hypergeometric orthogonal polynomials and their $q$-analogues}.
Springer Monographs in Mathematics.
(Berlin: Springer)


\bibitem{VZ}
Vinet L and Zhedanov A 2008
Generalized Bochner theorem: Characterization of the Askey-Wilson polynomials
{\it J. Comput. Appl. Math.} {\bf 211} 45-56
(arXiv:0712.0069)


\bibitem{KPe}
Kelley W G and Peterson  A C 2001
{\it Difference Equations.
An Introduction with Applications} 
2nd edn (Harcourt/Academic Press) Sec 3.2


\bibitem{DA}
Dimitrov D K and Van Assche W
2000 Lam\'e differential equations and electrostatics
{\it Proc. Amer. Math. Soc.} {\bf 128} 3621-8


\bibitem{Lsk}
Lesky P 1962
Die Charakterisierung der klassischen 
orthogonalen Polynome durch Sturm-Liouvillesche 
Differentialgleichungen
{\it Arch. Rat. Mech. Anal.} {\bf 10} 341-51


\bibitem{EG}
Eremenko A and Gabrielov A 2011
Elementary proof of the B. and M. Shapiro conjecture for rational functions.
Notions of positivity and the geometry of polynomials,
Trends Math. (Birkh\"auser/Basel: Springer) pp 167-78
(arXiv:math/0512370)


\bibitem{AMhs}
Moroz A 2015 
On uniqueness of Heine-Stieltjes polynomials for second order
finite-difference equations with Hahn operator, submitted for publication


\bibitem{H}
Hardy G H 1949 {\it Divergent series}
(London: Oxford Univ. Press)


\bibitem{AMcjm}
Moroz A 1990
Analytic continuation by means of the methods of divergent series
{\it Czech. J. Math.} {\bf 40} 200-12


\end{thebibliography}
\end{document}